\documentclass
[preprint,pre,10pt,twocolumn,superscriptaddress,notitlepage,showpacs,nofootinbib]{revtex4}%
\usepackage{amsfonts}
\usepackage{amsmath}
\usepackage{amssymb}
\usepackage{graphicx}%
\newtheorem{theorem}{Theorem}

\newenvironment{proof}[1][Proof]{\noindent\textbf{#1.} }{\ \rule{0.5em}{0.5em}}
\begin{document}
\title{Thermo-Statistical description of the Hamiltonian non extensive systems\\The reparametrization invariance}
\author{L. Velazquez}
\email{luisberis@geo.upr.edu.cu}
\affiliation{Departamento de F\'{\i}sica, Universidad de Pinar del Rio, Marti 270, Esq. 27
de Noviembre, Pinar del Rio, Cuba}
\author{F. Guzman}
\affiliation{Instituto Superior de Tecnologia y Ciencias Aplicadas, Quinta de los Molinos,
Plaza de la Revolucion, La Habana, Cuba}

\begin{abstract}
In the present paper we continue our reconsideration about the foundations for
a thermostatistical description of the called Hamiltonian nonextensive
systems. After reviewing the selfsimilarity concept and the necessary
conditions for the ensemble equivalence, we introduce the reparametrization
invariance of the microcanonical description as an internal symmetry
associated with the dynamical origin of this ensemble. Possibility of
developing a geometrical formulation of the thermodynamic formalism based on
this symmetry is discussed, with a consequent revision about the
classification of phase-transitions based on the concavity of the Boltzmann
entropy. The relevance of such conceptions are analyzed by considering the
called Antonov isothermal model.

\end{abstract}
\date{\today}
\maketitle

\section{Introduction}

As alsewhere discussed, the study of the non extensive systems constitutes an
interesting and fascinating challenger for the developments of the
Thermodynamics and Statistical Mechanics
\cite{bog,bec,sol,lyn,pie,pos,kon,tor,gro1,ato,kud,par,stil,sta1}. An
important feature of this kind of systems is the existence of long-range
correlations, whose origin delays on the existence of long-range interactions
among the constituents particles or the small (or mesoscopic) nature of the system.

We address in our previous paper \cite{vel.self} a reconsideration of the
foundations for a thermo-statistical description of the special case of the
nonextensive Hamiltonian systems. The main conclusions derived from this work
are the following: (1) The universality of the microscopic mechanism of
chaoticity supports the generic applicability of a thermostatistical
description with microcanonical basis for the nonintegrable many-body
Hamiltonian systems \cite{Kolmogorov,Arnold,Moser,PF1,PF2,arnold}, even for
those nonextensive systems with long-range interactions
\cite{pettini4793,pettini 51,pet,cohenG}, (2) Extensive conditions of the
traditional systems are not longer applicable for the nonextensive systems,
but such properties could be generalized by considering the scaling
selfsimilarity concept \cite{vel.self} which is considered in order to
establish an adequate thermo-statistical description in this context.

We will continue in the present paper our analysis of the foundations for a
thermo-statistical description of the nonextensive Hamiltonian systems. Our
aim now is to show how the scaling selfsimilarity concept could be used in
order to perform an appropriate thermodynamic formalism in the systems with
long-range interactions. As already discussed in our previous work, most of
the Hamiltonian systems with a practical interest exhibit exponential
selfsimilarity scaling laws \cite{vel.self}. This is the reason why the
present discussion will be focussed in this special class of the nonextensive
Hamiltonian systems. Interested reader could also review a similar approach to
the one developed in this work in ref. \cite{velTsal} for the case of the
potential selfsimilarity scaling laws, which seem to be associated with the
popular Tsallis nonextensive statistics \cite{tsal}.

\section{Reviewing selfsimilarity\label{rev}}

Selfsimilarity is just a general symmetry of the macroscopic description
exhibited under the scaling transformations of the system size in $\alpha$
times, $N\rightarrow N\left(  \alpha\right)  =\alpha N$, which manifests
itself in every macroscopic behavior of the system, even in the macroscopic
dynamics \cite{vel.self}. From the mathematical viewpoint, selfsimilarity
appears as a multiplicative uniparametric group of transformations $T_{\alpha
}$, $T_{\alpha}T_{\beta}=T_{\alpha\beta}$, which acts on those fundamental
macroscopic observables $I$ determining the microcanonical macroscopic state
of a given system, leading in this way to a scaling transformation of the
number of microscopic configurations $W=W\left(  I\right)  $ (microcanonical
accessible phase-space volume) whose form does not depend on the macroscopic
state and can be represented as follows:%

\begin{equation}
I\left(  \alpha\right)  =T_{\alpha}I\Rightarrow W\left[  \alpha\right]
=\mathcal{Y}\left\langle \alpha\mathcal{Y}^{-1}\left(  W\right)  \right\rangle
,
\end{equation}
where the function $\mathcal{Y}\left(  x\right)  $\ determines what we refer
as the selfsimilarity scaling law of the system. According to the framework
developed in the previous work \cite{vel.self}, the selfsimilarity scaling law
$\mathcal{Y}\left(  x\right)  $ determines the generalized Boltzmann Principle
(microcanonical entropy) and the specific statistical entropic compatible with
the selfsimilatity of a given system:%

\begin{equation}
S_{B}^{\mathcal{L}}=\mathcal{L}\left(  W\right)  \Leftrightarrow
S_{e}^{\mathcal{L}}=\sum_{k}p_{k}\mathcal{L}\left(  \frac{1}{p_{k}}\right)  ,
\end{equation}
where $\mathcal{L}\left(  x\right)  =\mathcal{Y}^{-1}\left(  x\right)  $. Such
definitions should be considered as the starting point in the generalization
of the traditional thermodynamic formalism.

Two remarkable cases of selfsimilarity scaling laws are the
\textit{exponential selfsimilarity}:%
\begin{equation}
\mathcal{Y}\left(  x\right)  =\exp\left(  x\right)  \Rightarrow S_{BG}%
=-\sum_{k}p_{k}\ln p_{k}, \label{exp}%
\end{equation}
which leads to the usual Boltzmann-Gibbs statistics, and the \textit{potential
selfsimilarity}:%
\begin{equation}
\mathcal{Y}\left(  x\right)  =\exp_{q}\left(  x\right)  \Rightarrow
S_{q}=-\sum_{k}p_{k}^{q}\ln_{q}p_{k},
\end{equation}
associated with the Tsallis nonextensive statistics \cite{tsal}, where
$\ln_{q}x=\exp_{q}^{-1}\left(  x\right)  =\left(  x^{1-q}-1\right)  /\left(
1-q\right)  $, is the \textit{q}-generalized logarithmic function. Among them,
the most extended in practical applications is the exponential selfsimilarity
scaling laws (\ref{exp}). Extensivity of traditional systems is just a special
case of this kind of scaling selfsimilarity, whose scaling selfsimilarity
transformations obey the well-kown form:%
\begin{equation}
\left.
\begin{array}
[c]{c}%
N\left(  \alpha\right)  =\alpha N\\
E\left(  \alpha\right)  =\alpha E\\
V\left(  \alpha\right)  =\alpha V
\end{array}
\right\}  \Rightarrow W\left(  \alpha\right)  =\exp\left(  \alpha\ln W\right)
. \label{extensivity}%
\end{equation}

While a generic Hamiltonian system of the form:%
\begin{equation}
H\left(  q,p\right)  =\sum_{ij}\frac{1}{2}a^{ij}p_{i}p_{j}+V\left(  q\right)
,
\end{equation}
with the total potential energy $V\left(  q\right)  $, $i,j=1,2,\ldots n$ and
$q=\left(  q^{1},q^{2},\ldots q^{n}\right)  $, exhibits an exponential growing
of the microcanonical accessible phase-space volume $W$ with the increasing of
the system degrees of freedom $n$ \cite{vel.self}, supporting in this way an
exponential growing of $W$ during the scaling operation $n\rightarrow n\left(
\alpha\right)  =\alpha n$, the selfsimilarity scaling transformations
$T_{\alpha}$ acting on the fundamental macroscopic observables (such as the
total energy $E$ or the systems volume $V$) are far to look like the extensive
case (\ref{extensivity}) in the long-range interacting systems.

A very simple example of system with nontrivial scaling transformations is
found for the selfgravitating gas of identical nonrelativistic particles with
a finite incompressible volume $v$ which interact among them by means of the
Newtonian gravity. Obviously, the particles density $\rho$ in this system
obeys the inequallity $\rho\leq\rho_{0}=1/v$. For very low energies and large
densities where $\rho\sim\rho_{0}$ the scaling transformations are
approximately given by:
\begin{equation}
N\left(  \alpha\right)  =\alpha N,~E\left(  \alpha\right)  \simeq\alpha
^{\frac{5}{3}}E,V\left(  \alpha\right)  \simeq\alpha V,
\end{equation}
while for very high energies and low densities where the particles size can be
dismissed since $\rho\ll\rho_{0}$, such scaling transformations can be
expressed asyntotically as follows:%

\begin{equation}
N\left(  \alpha\right)  =\alpha N,~E\left(  \alpha\right)  \simeq\alpha
^{\frac{7}{3}}E,V\left(  \alpha\right)  \simeq\alpha^{-1}V, \label{diluted}%
\end{equation}
results which have been obtained in refs.\cite{velastro,chava12} and it will
be shown again in the section \ref{astro}.

Generally speaking, the determination of the specific selfsimilarity scaling
transformations $T_{\alpha}$ could be a very difficult task because of there
is no methodology to establish them in a given application outside the
extensive context. As already suggested in the introductory section of our
previous paper \cite{vel.self}, systems with long-range interactions look like
the extensive systems in the neighborhood of the second-order phase
transitions due to the long-range correlations existing in this region. This
is the reason why we think that the well-known Renormalization Group methods
\cite{Gold} should play a crucial role in the determination of the scaling
selfsimilarity of a given nonextensive system.

\section{Thermodynamic formalism\label{ThF}}

As elsewhere discussed, the thermodynamic formalism is based on the
equivalence in the thermodynamic limit $N\rightarrow\infty$ between the
microcanonical ensemble $\hat{\omega}_{M}$ and the canonical ensemble
$\hat{\omega}_{BG}$ \cite{gallavotti}:
\begin{equation}
\hat{\omega}_{M}=\frac{1}{\Omega\left(  I,N\right)  }\delta\left\langle
I-\hat{I}_{N}\right\rangle ,~\hat{\omega}_{BG}=\frac{1}{Z\left(
\beta,N\right)  }\exp\left(  -\beta\hat{I}_{N}\right)  , \label{pdf}%
\end{equation}
derived from the consideration of the Maximal Entropy Principle \cite{jaynes}
by using the extensive entropy (\ref{exp}), where $\hat{I}_{N}$ are all those
macroscopic observables (integrals of motion) determining the macroscopic
state. Obviously, the thermodynamic limit $N\rightarrow\infty$ is equivalent
to carry out a scaling transformation $T_{\alpha}$ on the observables $I$ with
$\alpha\rightarrow\infty$.

The ensemble equivalence can be shown by considering the Laplace
transformation which relates their partition functions:%

\begin{equation}
Z\left(  \beta,N\right)  =\int e^{-\beta I}\Omega\left(  I,N\right)  dI.
\end{equation}
Introducing the Boltzmann entropy $S_{B}\left(  I,N\right)  =\ln W$, where
$W=\Omega\delta I$\footnote{Here $\delta I$ is a small volume constant
associated with the coarsed grained of the phase-space.}, and the Planck
thermodynamic potential $\mathcal{P}\left(  \beta\right)  =-\ln Z\left(
\beta\right)  $, the Laplace transformation could be rewritten as follows:%

\begin{equation}
\exp\left\langle -\mathcal{P}\left(  \beta,N\right)  \right\rangle =\int
\exp\left\langle -\left[  \beta I-S_{B}\left(  I,N\right)  \right]
\right\rangle \frac{dI}{\delta I}.
\end{equation}
We immediately recognize in the argument of the exponential function the
well-known Legendre transformation between the thermodynamic potentials:%

\begin{equation}
\mathcal{P}_{m}\left(  \beta_{m},N\right)  =\beta_{m}I-S_{B}\left(
I,N\right)  ,\text{ with }\beta_{m}=\frac{\partial S_{B}}{\partial I},
\label{Legendre}%
\end{equation}
where the subindex $m$ means that such quantities are associated with the
microcanonical description.

It can be shown by using the steepest descend method that the equivalence
between $\mathcal{P}_{m}\left(  \beta_{m},N\right)  $ and $\mathcal{P}\left(
\beta,N\right)  $ takes place in the thermodynamic limit when the following
conditions are applicable:

\begin{itemize}
\item[ \textbf{i)}] The macroscopic observables $I$ behave in a
\textit{extensive-like} way during the selfsimilarity scaling transformation
$T_{\alpha}$:
\begin{equation}
T_{\alpha}I=\alpha I\Rightarrow S_{B}\left(  T_{\alpha}I,\alpha N\right)
=\alpha S_{B}\left(  I,N\right)  . \label{ext-like}%
\end{equation}

\item[\textbf{ii)}] There is only one stationary point $i_{s}=I/N$ satisfying
the relation:%
\begin{equation}
\beta=\frac{\partial s_{B}\left(  i_{s}\right)  }{\partial i},
\end{equation}
where $s_{B}\left(  i\right)  =\frac{1}{N}S_{B}\left(  iN,N\right)  $ is the
entropy per particle.
\end{itemize}

The condition \textbf{i)} is only trivially satisfied by the extensive systems
when the microcanonical description is performed by using the called extensive
observables, i.e. total energy, electric charge, etc. However, extensivity
does not take place in systems with long-range interactions. In this context
extensivity should be substituted by the scaling selfsimilarity, but the
extensive-like character (\ref{ext-like}) of the macroscopic observables $I$
does not hold generally speaking.

The condition \textbf{ii)} demands the concavity of the Boltzmann entropy,
property which can not be everywhere satisfied. The existence of several
stationary points leads to a catastrophe in the Legendre transformation which
is related with the occurrence of first-order phase transitions at a given
value of $\beta$. Such anomaly appears in those regions where at least one
eigenvalue of the \textit{Hessian} of the Boltzmann entropy:%
\begin{equation}
k_{nm}=\frac{\partial^{2}s_{B}}{\partial i^{n}\partial i^{m}}, \label{hessian}%
\end{equation}
is positive. As already shown by Gross, every anomaly in the Hessian of the
Boltzmann entropy can be related with the existence of phase transitions,
extending in this way the identification of such critical phenomena is systems
outside the thermodynamic limit \cite{gro1}.

\section{Reparametrization invariance\label{rep}}

We will show in this section that the conditions \textbf{i)} and \textbf{ii)}
could be satisfied by considering the called reparametrization invariance of
the microcanonical description. \textquestiondown What is such
\textit{reparametrization invariance}? As already commented, the
microcanonical ensemble is just a dynamical ensemble, and therefore, every
symmetry of the microscopic dynamics is also relevant at the macroscopic level
in the microcanonical description. The reparametrization invariance is just a
symmetry with a dynamical origin.

\subsection{Mathematical definition and Physical meaning}

Let us consider now all those integrals of motion $\hat{I}\left(  X\right)
=\left\{  I^{1}\left(  X\right)  ,I^{2}\left(  X\right)  ,\ldots I^{n}\left(
X\right)  \right\}  $ determining the microcanonical thermostatistical
description of a given nonintegrable many-body Hamiltonian system. Each
admissible value of the above set integrals of motion $I=\left\{  I^{1}%
,I^{2},\ldots I^{n}\right\}  $ determine certain subset $\mathcal{S}_{p}$ of
the phase-space $\mathcal{X}$ :%
\begin{equation}
X\in\mathcal{S}_{p}\equiv\left\{  X\in\mathcal{X}\left\vert I^{k}\left(
X\right)  =I^{k},k=1,2,\ldots n\right.  \right\}  . \label{set}%
\end{equation}
in which is enclose the system microscopic dynamics in a given microcanonical
macroscopic state.

The totality of these admissible "points" $\left\{  I\right\}  $ defines\ a
subset $\mathcal{R}_{I}$ of the n-dimensional Euclidean space $\mathcal{R}%
^{n}$, while the totality of the corresponding phase-space subsets $\left\{
\mathcal{S}_{p}\right\}  $\ defines certain partition $\Im\left(
\mathcal{X}\right)  $ of the phase-space $\mathcal{X}$:
\begin{equation}
\Im\left(  \mathcal{X}\right)  =\left\{  \mathcal{S}_{p}\subset\mathcal{X}%
\left\vert ~%
%TCIMACRO{\dbigcup _{p}}%
%BeginExpansion
{\displaystyle\bigcup_{p}}
%EndExpansion
\mathcal{S}_{p}=\mathcal{X}~;~\mathcal{S}_{p}\cap\mathcal{S}_{q}%
=\varnothing\right.  \right\}  . \label{partition}%
\end{equation}
The above definitions (\ref{set}) and (\ref{partition}) allow the existence of
certain bijective map $\psi_{I}$ between the elements of $\Im\left(
\mathcal{X}\right)  $ (subsets $\mathcal{S}_{p}\subset\mathcal{X}$) and the
elements of $\mathcal{R}_{I}$ (points $I\in\mathcal{R}^{n}$)%

\begin{align}
\psi_{I}  &  :\Im\left(  \mathcal{X}\right)  \rightarrow\mathcal{R}_{I}%
\equiv\left\{  \forall\mathcal{S}_{p}\in\Im\left(  \mathcal{X}\right)
~\exists I\in\mathcal{R}_{I}\subset\mathcal{R}^{n}\right. \nonumber\\
&  \left\vert \forall X\in\mathcal{S}_{p}\Rightarrow I^{k}\left(  X\right)
=I^{k},k=1,2,\ldots n\right\}  .
\end{align}
The bijective character of this map provides to the partition $\Im\left(
\mathcal{X}\right)  $ the same topological features of certain subset
$\mathcal{R}_{I}$ of the n-dimensional Euclidian space $\mathcal{R}^{n}$. Now
on we refer the partition $\Im\left(  \mathcal{X}\right)  $ as the
\textit{abstract space of the integrals of motions}, and it will be denoted
simply by $\Im$. We say that the map $\psi_{I}$ defines a n-dimensional
Euclidean coordinate representation $\mathcal{R}_{I}$ of the\ abstract space
$\Im$.

The\ coordinate representation $\mathcal{R}_{I}$ of $\Im$ is not unique. Let
$\mathcal{R}_{I}$ and $\mathcal{R}_{\varphi}$ be two diffeomorphic subsets of
the n-dimensional Euclidean space $\mathcal{R}^{n}$, that is, there exist a
diffeomorphic map:%

\begin{equation}
\varphi:\mathcal{R}_{I}\rightarrow\mathcal{R}_{\varphi}\equiv\left\{  \forall
I\in\mathcal{R}_{I}~\exists\varphi\in\mathcal{R}_{\varphi}\left\vert
\det\left(  \frac{\partial\varphi^{j}}{\partial I^{k}}\right)  \not =0\right.
\right\}  , \label{diffeomorphic}%
\end{equation}
among their elements. The terms "diffeomorphic map" means that the
n-dimensional Euclidean subsets $\mathcal{R}_{I}$ and $\mathcal{R}_{\varphi}$
posses the same topological and diffeomorphic (differential) structure. The
map $\psi_{\varphi}=\psi_{I}o\varphi^{-1}$ defines a new bijective application
among the elements of the space $\Im$ and the points $\varphi\in
\mathcal{R}_{\varphi}$, $\psi_{\varphi}:$ $\Im\rightarrow\mathcal{R}_{\varphi
}$, and consequently, $\psi_{\varphi}$ is also another n-dimensional Euclidean
coordinate representation of $\Im$. The coordinate representations
$\mathcal{R}_{I}$ and $\mathcal{R}_{\varphi}$ are equivalent because of they
represent the same abstract space $\Im$.

Generally speaking, a reparametrization change is carried out when we move
from the Euclidean representation $\mathcal{R}_{I}$ towards an equivalent
Euclidean representation $\mathcal{R}_{\varphi}$. The totality of the
diffeomorphic maps (\ref{diffeomorphic}) constitutes certain group of
transformations called the \textit{group of diffeomorphisms} of the space
$\Im$, $Diff\left(  \Im\right)  $, which is the maximal symmetry that a
geometrical theory could exhibit.

Every reparametrization change (\ref{diffeomorphic}) also induces a
reparametrization of the integrals of motions $\varphi\Rightarrow\varphi_{X}%
$:
\begin{equation}
\varphi_{X}:\hat{I}\left(  X\right)  \rightarrow\hat{\varphi}\left(  X\right)
\equiv\left\{  \varphi^{1}\left\langle \hat{I}\left(  X\right)  \right\rangle
,\ldots\varphi^{n}\left\langle \hat{I}\left(  X\right)  \right\rangle
\right\}  , \label{changeX}%
\end{equation}
Since $\hat{I}\left(  X\right)  $ are integrals of motions, every $\varphi
^{k}\left\langle I\left(  X\right)  \right\rangle \in$ $\hat{\varphi}\left(
X\right)  $ will be also a integral of motion. Due to the bijective character
of the reparametrization change $\varphi:\mathcal{R}_{I}\rightarrow
\mathcal{R}_{\varphi}$, the new set of integrals of motion $\hat{\varphi
}\left(  X\right)  $ generates the same partition $\Im\left(  \mathcal{X}%
\right)  $ of the phase-space. This is the reason why the sets $\hat{I}\left(
X\right)  $ and $\hat{\varphi}\left(  X\right)  $ can be considered as
\textit{equivalent representations} of the relevant integrals of motion
determining the microcanonical description.

Since the totality of the phase-space points belonging to a given subset
$\mathcal{S}_{p}\in$ $\Im\left(  \mathcal{X}\right)  $ represents the same
microcanonical macroscopic state, and the diffeomorphic and topological
structure of the abstract space $\Im$ does not depend on the reparametrization
changes, it is not difficult to understand that the microcanonical description
is reparametrization invariant.

\begin{theorem}
\label{inv_w}The microcanonical ensemble $\hat{\omega}_{M}$ (\ref{pdf}) is
invariant under every reparametrization change $\varphi:\mathcal{R}_{I}$
$\rightarrow\mathcal{R}_{\varphi}$.
\end{theorem}

\begin{proof}
Such invariance is straightforward derived from the following identity of the
Dirac delta function:%
\begin{equation}
\delta\left\langle \varphi\left(  I\right)  -\varphi\left(  \hat{I}%
_{N}\right)  \right\rangle =\left\vert \frac{\partial\varphi}{\partial
I}\right\vert ^{-1}\delta\left\langle I-\hat{I}_{N}\right\rangle ,
\end{equation}
which leads to the following transformation rule for the partition function
$\Omega$:%
\begin{equation}
\Omega\left(  \varphi;N\right)  =\left\vert \frac{\partial\varphi}{\partial
I}\right\vert ^{-1}\Omega\left(  I;N\right)  ,
\end{equation}
and consequently:%
\begin{align}
\frac{1}{\Omega\left(  \varphi,N\right)  }\delta\left\langle \varphi
-\hat{\varphi}_{N}\right\rangle  &  =\frac{1}{\Omega\left(  I,N\right)
}\delta\left\langle I-\hat{I}_{N}\right\rangle ,\nonumber\\
\hat{\omega}_{M}\left(  \varphi,N\right)   &  =\hat{\omega}_{M}\left(
I,N\right)  . \label{invariance}%
\end{align}
We have considered here the implicit dependence of integrals of motion on the
system size $N$.
\end{proof}

A corollary of the Theorem \ref{inv_w} is that the Physics derived from the
microcanonical description is reparametrization invariant since the
expectation values of every macroscopic observable $\hat{O}$ derived from the
microcanonical distribution function $\hat{\omega}_{M}$ exhibits this kind of
symmetry:%
\begin{equation}
\bar{O}=\int\hat{O}\hat{\omega}_{M}dX\Rightarrow\bar{O}\left(  \varphi
,N\right)  =\bar{O}\left(  I,N\right)  ,
\end{equation}
relation which is straightforwardly followed by using the identity
(\ref{invariance}).

The reparametrization invariance does not introduce anything new in the
macroscopic description except the possibility of describing the macroscopic
state by using any representation $\mathcal{R}_{\varphi}$ equivalent to the
original representation $\mathcal{R}_{I}$, a situation analogue to describe
the physical space $\mathcal{R}^{3}$ by using a Cartesian coordinates $\left(
x,y,z\right)  $ or spherical coordinates $\left(  r,\theta,\varphi\right)  $.

Notice that the diffeomorphic reparametrization changes demand the analyticity
of the integrals of motions, which is a necessary condition for a particular
integral of motion $\hat{I}^{k}$ be relevant in the macroscopic description.
On the other hand, the integrals of motions $\hat{I}$ are derived from their
commutativity with the Hamiltonian $\hat{H}$. Since $\hat{H}$ is one of those
integrals of motion determining the macroscopic state (total energy), the
Hamiltonian loses its identity during the reparametrization changes
(\ref{changeX}). The only way to preserve the condition $\left\{  \hat{H}%
,\hat{I}\right\}  =0$ during the reparametrization changes is\ by demanding
the commutativity among all those integrals of motion determining the
macroscopic state $\left\{  \hat{I}^{k},\hat{I}^{l}\right\}  \equiv0$:
\begin{equation}
\left\{  \hat{\varphi}^{i},\hat{\varphi}^{j}\right\}  =\sum_{kl}\frac
{\partial\varphi^{i}}{\partial I^{k}}\frac{\partial\varphi^{j}}{\partial
I^{l}}\left\{  \hat{I}^{k},\hat{I}^{l}\right\}  \equiv0.
\end{equation}
Interestingly, this is the necessary condition for the simultaneous
determination of physical observables in the Quantum Mechanics, and therefore,
reparametrization invariance is also consistent in this level.

\subsection{Some geometrical aspects}

Reparametrization invariance is a symmetry of the microcanonical description
which allows us to perform a geometrical description in this framework, a
covariant formulation of the Thermostatistics. Geometrical formulations of the
Thermostatistics are very well-known in Physics. Some recently proposed
geometrical formulations are given in refs. \cite{gro1} and
\cite{rupper,topH2}.

A very important question is to determine the transformation rules of the
physical observables and the thermodynamical function under the
reparametrization changes. Now on the Einstein summation convention will be
assumed. As already shown, the microcanonical distribution function and the
physical observables behave as scalar functions during the reparametrization
changes:%
\begin{equation}
\hat{\omega}_{M}\left(  \varphi,N\right)  =\hat{\omega}_{M}\left(  I,N\right)
,~\bar{O}\left(  \varphi,N\right)  =\bar{O}\left(  I,N\right)  .
\end{equation}
The transformation rule microcanonical partition function $\Omega$ corresponds
with the transformation rule of a scalar tensorial density:%
\begin{equation}
\Omega\left(  \varphi,N\right)  =\left\vert \frac{\partial\varphi}{\partial
I}\right\vert ^{-1}\Omega\left(  I,N\right)  .
\end{equation}
The first derivatives of a given physical observable $\bar{O}$ behave as the
components of a covariant vector:%
\begin{equation}
\frac{\partial\bar{O}}{\partial\varphi^{m}}=\frac{\partial I^{k}}%
{\partial\varphi^{m}}\frac{\partial\bar{O}}{\partial I^{k}}\sim\upsilon
_{m}=\frac{\partial I^{k}}{\partial\varphi^{m}}\upsilon_{k}.
\label{rule_vector}%
\end{equation}
The transformation rule of the second-order tensors is given by:%

\begin{equation}
A_{mn}=\frac{\partial I^{k}}{\partial\varphi^{m}}\frac{\partial I^{l}%
}{\partial\varphi^{n}}A_{kl}. \label{rule_tensor}%
\end{equation}
However, such covariant second-order tensor cannot be obtained by
differentiation without the introduction of a Riemannian structure, that is,
without introducing a Riemannian metric.

The microcanonical partition function $\Omega$ allows us the introduction of
the invariant measure $d\mu=\Omega dI$. Taking into account such invariant
measure, the integration of a given physical observable $\bar{O}$ defined on
the space $\Im$ (microcanonical ensemble) is directly related with the
phase-space integration:%

\begin{equation}%
%TCIMACRO{\dint \limits_{\Sigma}}%
%BeginExpansion
{\displaystyle\int\limits_{\Sigma}}
%EndExpansion
\bar{O}d\mu\equiv%
%TCIMACRO{\dint \limits_{\Sigma\left(  \mathcal{X}\right)  }}%
%BeginExpansion
{\displaystyle\int\limits_{\Sigma\left(  \mathcal{X}\right)  }}
%EndExpansion
O\left(  X\right)  dX,
\end{equation}
where $\Sigma$ is a subset of the space $\Im$, and $\Sigma\left(
\mathcal{X}\right)  $ is its corresponding image in the phase-space.

Strictly speaking, the Boltzmann entropy $S_{B}=\ln W$ is not an scalar
function. The microcanonical accessible volume $W$ is just the invariant
measure of a small neighborhood $\Sigma$ of the interest point obtained after
by performing a coarsed grained partition of the space $\Im$:%

\begin{equation}
W=\mu\left(  \Sigma\right)  =%
%TCIMACRO{\dint \limits_{\Sigma}}%
%BeginExpansion
{\displaystyle\int\limits_{\Sigma}}
%EndExpansion
d\mu.
\end{equation}
In applications $W$ is taken approximately by $W\simeq\Omega\delta I$, where
the volume $\delta I$ of the coarsed grain subset $\Sigma$ is regarded as a
very small constant. The small constant $\delta I$ becomes unimportant when
$N$ is very large, that is, with the imposition of the thermodynamic limit,
and consequently, the Boltzmann entropy could be considered as a scalar
function in this limit.

\subsection{Consequences of this dynamical symmetry on the thermodynamic
formalism\label{conseq}}

The microcanonical ensemble provides a full characterization of the
macroscopic properties of a given isolate Hamiltonian system in thermodynamic
equilibrium. The imposition of the thermodynamic limit $N\rightarrow\infty
$\ leads under certain conditions to the equivalence of the microcanonical
description with certain simplified characterization: the macroscopic
description performed by the canonical ensemble. As already discussed, such
canonical ensemble is determined from the relevant statistical entropy derived
from the specific scaling selfsimilarity exhibited by the interest system,
i.e.: the exponential selfsimilarity is associated with the
Shannon-Boltzmann-Gibbs extensive entropy (\ref{exp}), which naturally leads
to the called Boltzmann-Gibbs distributions (\ref{pdf}).

It is very easy to verify that the canonical distribution function (\ref{pdf})
does not exhibit the original reparametrization invariance of the
microcanonical ensemble. This means that the Physics in the canonical ensemble
depends on the coordinate representation $\mathcal{R}_{I}$ of the abstract
space $\Im$ of the relevant integrals of motion used for perfoming the
macroscopic characterization of a given system. Therefore, the thermodynamic
formalism could not satisfy this kind of symmetry, and consequently, the
ordering information derived from its consideration is not reparametrization invariant.

As already discussed in section \ref{ThF}, the equivalence between the
microcanonical and the canonical descriptions demands the fulfillment the
conditions \textbf{i)} and \textbf{ii)}, and therefore, such ensemble
equivalence could not be performed by using an arbitrary representation.
However, the existence of the reparametrization invariance of the
microcanonical descriptions plays a crucial role in arriving to a well-defined
thermodynamic formalism.

The extensive-like behavior of the integrals of motion under the scaling
transformations is simply satisfied by chosing an appropriate coordinate
representation $\mathcal{R}_{\varphi}$\ of the space $\Im$\ exhibiting this
kind of behavior under the scaling transformations. There are some trivial
examples in which this demand is very easy to accomplish. Considering the
example illustrated at the end of section \ref{rev}, the total energy of the
selfgravitating gas of identical point particles interacting by means of the
Newtonian gravity grows with the scaling following a $\alpha^{\frac{7}{3}}$
power law \cite{velastro}, and therefore, an appropriate representation with
an extensive-like behavior could be $\varphi=N\varphi\left(  E/E_{0}\right)
$, where $E_{0}=GM^{2}/R$ is the characteristic constant energy which
evidently follows the same $\alpha^{\frac{7}{3}}$ power law, being
$\varphi\left(  x\right)  $ a bijective function.

The reader may notice that such procedure differs from the called Kac argument
\cite{kac} because of the total energy is not arbitrarily forced to be
extensive, since its scaling behavior is determined from the system
selfsimilarity. We only demand the extensivity in the \textit{coordinate
representation} of the relevant integrals of motion in order to ensure the
ensemble equivalence in the thermodynamic limit. The main difficulty in
chosing an appropriate representation of the relevant integrals of motion is
that the specific selfsimilarity scaling transformation $T_{\alpha}$ acting on
the space $\Im$ is \textit{a priori} unknown and have to be determined.

As already shown, the transition from the microcanonical towards the canonical
description with the imposition of the thermodynamic limit and the consequent
establishment of the system selfsimilarity leads to a breakdown of the
original reparametrization symmetry. However, \textquestiondown is the
reparametrization invariance completely broken during this transition? The
answer is \textit{no}. This fact is very easy to understand by considering the
above example about the selfgravitating gas: anyone bijective function
$\varphi=N\varphi\left(  E/E_{0}\right)  $ could be used with the purpose of
describing the macroscopic characterization, and therefore, there is a
complete set of admissible coordinate representations $\mathcal{C}=\left\{
\mathcal{R}_{\varphi}\right\}  $ exhibiting an extensive-like behavior during
the scaling transformations.

Let $\mathcal{R}_{I}$ and $\mathcal{R}_{\varphi}$ be two coordinate
representations belonging to the set of admissible representations
$\mathcal{C}$ and $\varphi=\psi_{\varphi}o\psi_{I}^{-1}:\mathcal{R}%
_{I}\rightarrow\mathcal{R}_{\varphi}$, the reparametrization change between
them. Since the scaling transformations for every coordinate in both
representations are given by $T_{\alpha}^{k}\left(  I\right)  =\alpha I^{k}$
and $T_{\alpha}^{m}\left(  \varphi\right)  =\alpha\varphi^{m}$, the
reparametrization change $\varphi\left(  I\right)  $ is a \textit{homogeneous
map} satisfying the relations:%
\begin{equation}
\varphi^{m}\left(  \alpha I\right)  =\alpha\varphi^{m}\left(  I\right)
\Leftrightarrow\varphi^{m}\left(  I\right)  =I^{k}\frac{\partial\varphi
^{m}\left(  I\right)  }{\partial I^{k}}, \label{map-hom}%
\end{equation}
for $m=1,2,\ldots n$.

The complete set of reparametrization changes acting in the set of admissible
representations $\mathcal{C}$ constitutes the \textit{Group of Homogeneous
transformations} $\mathcal{H}$, which is a subgroup of the Group of
diffeormorphisms $Diff\left(  \Im\right)  $ already commented. The symmetry
associated with the transformations of the Homogeneous group $\mathcal{H}$ is
just the residual symmetry of the original reparametrization invariance of the
microcanonical description.

\begin{theorem}
The thermodynamic potential $\mathcal{P}_{m}$ obtained from the Legendre
transformation (\ref{Legendre}) is $\mathcal{H}$-invariant:%
\begin{equation}
\mathcal{P}_{m}=I^{k}\frac{\partial S_{B}\left(  I\right)  }{\partial I^{k}%
}-S_{B}\left(  I\right)  =\varphi^{m}\frac{\partial S_{B}\left(
\varphi\right)  }{\partial\varphi^{m}}-S_{B}\left(  \varphi\right)  .
\end{equation}

\end{theorem}

\begin{proof}
The demostration is straightforwardly followed by performing a variable change
in the partial derivative of the Boltzmann entropy:%
\begin{equation}
I^{k}\frac{\partial S_{B}}{\partial I^{k}}=I^{k}\left(  \frac{\partial
\varphi^{m}}{\partial I^{k}}\frac{\partial S_{B}}{\partial\varphi^{m}}\right)
=\left(  I^{k}\frac{\partial\varphi^{m}}{\partial I^{k}}\right)
\frac{\partial S_{B}}{\partial\varphi^{m}},
\end{equation}
and considering the homogeneous character of the reparametrization change
(\ref{map-hom}):%
\begin{equation}
I^{k}\frac{\partial S_{B}}{\partial I^{k}}\equiv\varphi^{m}\frac{\partial
S_{B}}{\partial\varphi^{m}}.
\end{equation}

\end{proof}

We have shown in this way that the "microcanonical" thermodynamic functions
$S_{B}$ and $\mathcal{P}_{m}$, as well as their first derivatives exhibit a
covariant behavior under the reparametrizations changes of the Homogeneous
group $\mathcal{H}$. However, it does not mean that the thermodynamic
formalism exhibits this kind of symmetry because of the conditions
\textbf{ii)} is necessary to satisfy also in order to ensure the ensemble
equivalence. The fulfillment of this condition could be perturbed because of
the reparametrization changes can \textit{affect the concavity} of the
Boltzmann entropy. Let us to illustrate this by considering the following example.

Let $\varphi$ be a map between two seminfinite Euclidean lines, $\varphi
:\mathcal{R}^{+}\rightarrow\mathcal{R}^{+}$, which in the representation
$\mathcal{R}_{x}$ is given by the concave function $\varphi\left(  x\right)
=\sqrt{x}$ (where $x>0$). Let us now consider a reparametrization change
$\psi:\mathcal{R}_{x}\rightarrow\mathcal{R}_{y}$ given by $y=x^{\frac{1}{4}}$
(which is evidently bijective). The map $\varphi$ in the new representation
$\mathcal{R}_{y}$ is given by the function $\varphi\left(  y\right)  =y^{2}$
(where $y>0$), which clearly is a convex function.

The above trivial example teaches us that the concavity of the Boltzmann
entropy depends on the representation $\mathcal{R}_{I}\in\mathcal{C}$ used to
perform the canonical description, and therefore, ensemble equivalence (or
inequivalence) depends on the coordinate representation of the space $\Im$.
This is a very interesting point since the ensemble inequivalence is a
signature of the first-order phase transitions.

According to the Microcanonical Thermostatistics of Gross \cite{gro1}, the
Hessian of the Boltzmann entropy (\ref{hessian})\ gives a complete
characterization of the phase-transitions in the microcanonical ensemble.
However, in spite of the aparent similarity, the components of the Hessian are
not the components of a covariant tensor of second-order because of they obey
the transformation rule:%
\begin{equation}
\frac{\partial^{2}S_{B}}{\partial\varphi^{m}\partial\varphi^{n}}%
=\frac{\partial I^{k}}{\partial\varphi^{m}}\frac{\partial I^{l}}%
{\partial\varphi^{n}}\frac{\partial^{2}S}{\partial I^{k}\partial I^{l}}%
+\frac{\partial^{2}I^{k}}{\partial\varphi^{m}\partial\varphi^{n}}%
\frac{\partial S}{\partial I^{k}},
\end{equation}
while the correct transformation rule is given by (\ref{rule_tensor}). This
means that such classification of phase-transitions based on the concavity of
the entropy is not reparametrization invariant. Therefore, the coordinate
representation of the space $\Im$ should be specified in order to precise the
existence of first-order phase transitions. How such ambiguity in the
first-order phase-transition existence could be physically acceptable?

The question is that the coordinate representation $\mathcal{R}_{I}$ of the
space $\Im$ is determined from the external constrains imposed to the system.
This idea is very easy to understand analysing the case of the extensive
systems. Ordinarily, canonical ensemble $\omega_{BG}=Z^{-1}\left(
\beta\right)  \exp\left(  -\beta E\right)  $ is experimentally implemented by
putting the interest system in a thermal contact with a heat bath. This
experimental arrangement not only fix the system temperature $T=\beta^{-1}$,
but also the coordinate representation by using the system energy $E$. Thus,
the existence of first-order phase transition is not necessarily an intrinsic
characteristic of a given system because of it depends on the external
conditions which have been imposed.

What happen when the ensemble inequivalence or any other anomaly is present at
a given point $p\in$ $\Im$ without matter the coordinate representation
$\mathcal{R}_{I}\in\mathcal{C}$ used in the macroscopic description? Such
anomalies are not representation-dependent, that is, they do not depend on the
external conditions imposed to the system. These kind of anomalies are
intrinsic characteristic of the system, and consequently, they have a
\textit{topological character}. A anomaly of this kind reflects the ocurrence
of considerable changes in the microscopic picture of an isolate Hamiltonian
system, i.e.: an ergodicity breaking \cite{Gold}.

This idea brings us the recently proposed hypothesis about the topological
origin of the phase-transitions \cite{topH2}. According to the ideas developed
by these authors, the origin of the chaoticity of the microscopic dynamics and
the\ origin of the phase transitions at the macroscopic level delay on the
geometrical and topological structure of the configurational space. We think
that such abrupt topological changes in the configurational space could be
took place when they are associated with topological anomalies at the
macrocopic level, that is, when such anomalies are not dependent of the
coordinate representation of the space $\Im$. The non correspondence between
topological changes and phase transitions observed in ref.\cite{ana} might be
explained due to the non topological character of such phase-transitions. The
following theorem establishes the non existence of topological ensemble
inequivalences in those systems controlling by only one scaling invariant
parameter $\varepsilon$, $T_{\alpha}\left(  \varepsilon\right)  =\varepsilon$.

\begin{theorem}
\label{existence}Let $s\left(  \varepsilon\right)  $ be the entropy per
particle of a given system which is controlled by only one scaling invariant
variable $\varepsilon$. Let us also suppose that the second derivative of
$s\left(  \varepsilon\right)  $ exists, being $s\left(  \varepsilon\right)  $
a convex function in the interval $\varepsilon_{1}<\varepsilon<\varepsilon
_{2}$, and a concave function elsewhere. There is a reparametrization change
$\varphi:\mathcal{R}_{\varepsilon}\rightarrow\mathcal{R}_{\varphi}$ in which
the entropy $s$\ becomes a concave function.
\end{theorem}

\begin{proof}
The transformation rule of the second derivative of the entropy during the
reparametrization change $\varphi:\mathcal{R}_{\varepsilon}\rightarrow
\mathcal{R}_{\varphi}$ is given by:
\begin{equation}
\frac{\partial^{2}s}{\partial\varphi^{2}}=\left(  \frac{\partial\varepsilon
}{\partial\varphi}\right)  ^{2}\frac{\partial^{2}s}{\partial\varepsilon^{2}%
}+\frac{\partial^{2}\varepsilon}{\partial\varphi^{2}}\frac{\partial
s}{\partial\varepsilon}.
\end{equation}
Considering the auxiliary function $a\left(  \varepsilon\right)  $:
\begin{equation}
\frac{\partial^{2}s}{\partial\varphi^{2}}=-\left(  \frac{\partial\varepsilon
}{\partial\varphi}\right)  ^{2}a\left(  \varepsilon\right)  ,
\end{equation}
the first derivative of the bijective map $\varphi\left(  \varepsilon\right)
$ could be rewritten as follows:
\begin{equation}
\frac{\partial\varphi\left(  \varepsilon\right)  }{\partial\varepsilon}%
=C\beta\left(  \varepsilon\right)  \exp\left(  \int\frac{a\left(
\varepsilon\right)  d\varepsilon}{\beta\left(  \varepsilon\right)  }\right)  ,
\end{equation}
where $C$ is certain positive integration constant and $\beta\left(
\varepsilon\right)  =\partial s\left(  \varepsilon\right)  /\partial
\varepsilon$. The entropy $s$ will be a concave function in the new coordinate
representation $\mathcal{R}_{\varphi}$ whenever $a\left(  \varepsilon\right)
>0$. This demand is very easy to satisfy by considering:%
\begin{equation}
a\left(  \varepsilon\right)  =\left\{
\begin{array}
[c]{cc}%
\partial^{2}s\left(  \varepsilon\right)  /\partial\varepsilon^{2} &
\text{wherever }s\left(  \varepsilon\right)  \text{ be convex}\\
-\partial^{2}s\left(  \varepsilon\right)  /\partial\varepsilon^{2} &
\text{wherever }s\left(  \varepsilon\right)  \text{ be concave}%
\end{array}
\right.  ,
\end{equation}
which directly leads to the following bijective map:%
\begin{equation}
\frac{\partial\varphi\left(  \varepsilon\right)  }{\partial\varepsilon
}=\left\{
\begin{array}
[c]{cc}%
C_{1} & \varepsilon<\varepsilon_{1}\\
C_{2}\beta^{2}\left(  \varepsilon\right)  & \varepsilon_{1}<\varepsilon
<\varepsilon_{2}\\
C_{3} & \varepsilon>\varepsilon_{2}%
\end{array}
\right.  ,
\end{equation}
where we have considered here the possibility of introducing a different
values for integration constant $C$ in different regions. By choosing
$C_{1}=1$, $C_{2}=1/\beta^{2}\left(  \varepsilon_{1}\right)  $ and
$C_{3}=\beta^{2}\left(  \varepsilon_{2}\right)  /\beta^{2}\left(
\varepsilon_{1}\right)  $, we not only ensure the concavity of the entropy,
but also the continuity of the first and second derivatives of the bijective
map $\varphi\left(  \varepsilon\right)  $.
\end{proof}

The analysis about the existence of topological ensemble inequivalences for
two or more controlling variables is an open problem. While topological
ensemble inequivalence are not present in the unidimensional case, they are
not the only topological anomalies presents in this context. The existence of
a discontinuity in the first derivative of the entropy in the thermodynamic
limit is a example of anomaly which could not be avoided by a
reparametrization change $\varphi\left(  \varepsilon\right)  $ with a continue
first derivative $\partial\varphi\left(  \varepsilon\right)  /\partial
\varepsilon$. Such anomalies have been recently reported in the astrophysical
context, and they are usually classified as \textit{microcanonical
phase-transitions} \cite{chava.micro}. Existence of singularities in the
second derivatives of the entropy are also unavoidable by reparametrizations,
which are related with the well-known second-order phase transitions appearing
due to the existence of an spontaneous symmetry breaking at the microscopic
level \cite{Gold}. Interestingly, the above topological anomalies appear due
to an ergodicity breaking in\ underlying the microscopic dynamics
\cite{Gold,chava.micro}.

\section{Astrophysical system\label{astro}}

Let us to illustrate the relevance of the scaling selfsimilarity and
reparametrization invariance by considering the microcanonical description of
a nontrivial nonextensive Hamiltonian system: the selfgravitating gas of
identical nonrelativistic point particles interacting throughout the Newtonian gravity:%

\begin{equation}
H_{N}=K_{N}+V_{N}=\sum_{i}\frac{1}{2m}\mathbf{p}_{i}^{2}-\frac{1}{2}Gm^{2}%
\sum_{\left\langle i\not =j\right\rangle }\frac{1}{\left\vert \mathbf{r}%
_{i}-\mathbf{r}_{j}\right\vert }.
\end{equation}
The long-range singularity of the gravitational potential will be avoided by
enclosing the system in a spherical rigid container of radio $R$. The
short-range singularity will be regularized by considering a mean-field
approximation. The above conditions leads to the well-known isothermal model
of Antonov \cite{antonov}.

After the integration of the linear momentum variables $\mathbf{p}$'s, the
accessibe phase-space volume can be expressed by:%
\[
W=\delta\varepsilon\frac{1}{N!}\left(  \frac{2m\pi}{h^{2}}\right)  ^{\frac
{3}{2}N}\int d^{3N}r\frac{1}{\Gamma\left(  \frac{3}{2}N\right)  }\left\langle
E-V_{N}\left\{  r\right\}  \right\rangle ^{\frac{3}{2}N-1},
\]
where $\left\{  r\right\}  =\left(  \mathbf{r}_{1},\mathbf{r}_{2}%
,\ldots\mathbf{r}_{n}\right)  $, $d^{3N}r=\prod_{k}d^{3}\mathbf{r}_{k}$, and
$\delta\varepsilon$ a very small energy constant. Introducing the
characteristics units for the linear dimension $l_{0}=R$, total mass $M=Nm$,
and total energy $E_{0}=GM^{2}/R$, and considering a very large system size
$N$ in order to consider the Stirling approximation $\Gamma\left(  x\right)
\simeq\left(  x/e\right)  ^{x}$, the above expression could be rewritten as
follows:%
\begin{equation}
\frac{1}{N}\ln W\simeq\frac{3}{2}\ln\left\langle C_{0}RN^{\frac{1}{3}%
}\right\rangle +\ln\omega\left(  U\right)  , \label{antonov_entropy}%
\end{equation}
where $C_{0}=4\pi Gm^{3}/3e^{\frac{5}{3}}h^{2}$ and $\omega\left(  U\right)  $
is given by:%
\begin{equation}
\ln\omega\left(  U\right)  =\lim_{N\rightarrow\infty}\frac{1}{N}\ln\left(
\int d^{3N}\xi\left\langle U-V_{N}\left\{  \xi\right\}  \right\rangle
^{\frac{3}{2}N}\right)  , \label{w_antonov}%
\end{equation}
being $U=E/E_{0}$ and $\xi=r/l_{0}$. The entropy per particle
(\ref{antonov_entropy}) is scaling invariant under the following scaling
selfsimilarity transformations:%

\begin{equation}
N\left(  \alpha\right)  =\alpha N,R\left(  \alpha\right)  =\alpha^{-\frac
{1}{3}}R\Rightarrow E\left(  \alpha\right)  =\alpha^{\frac{7}{3}}E\text{,}%
\end{equation}
result supporting the thermodynamic limit:%
\begin{equation}
N\rightarrow\infty\text{, keeping fixed }\frac{E}{N^{\frac{7}{3}}}\text{ and
}RN^{\frac{1}{3}}%
\end{equation}
already obtained in refs.\cite{velastro,chava12}. Now on $s_{0}\left(
U\right)  =\ln\omega\left(  U\right)  $ will be referred simply as the system entropy.

The progresive calculation of (\ref{w_antonov}) is carried out by considering
a mean-field approximation and them using the steepest descend method whose
details will be omitted here in the sake of brevity. Such procedure leads to
the following \textit{min}-\textit{max} problem:%
\begin{equation}
s_{0}\left(  U\right)  \simeq\min_{\mu}\max_{\rho}\left\{  \frac{3}{2}%
\ln\left\langle U-V\left[  \rho\right]  \right\rangle +s\left[  \rho\right]
+\mu\left\langle 1-N\left[  \rho\right]  \right\rangle \right\}  ,
\label{extreme}%
\end{equation}
where the functionals $V\left[  \rho\right]  $ (potential energy), $s\left[
\rho\right]  $ (local entropy) and $N\left[  \rho\right]  $ (particle number)
are given by:%
\begin{align}
V\left[  \rho\right]   &  =-\frac{1}{2}\int d^{3}\mathbf{\xi}d^{3}\mathbf{\xi
}^{\prime}\frac{\rho\left(  \mathbf{\xi}\right)  \rho\left(  \mathbf{\xi
}^{\prime}\right)  }{\left\vert \mathbf{\xi}-\mathbf{\xi}^{\prime}\right\vert
},~N\left[  \rho\right]  =\int d^{3}\mathbf{\xi}~\rho\left(  \mathbf{\xi
}\right)  ,\nonumber\\
&
\begin{array}
[c]{c}%
s\left[  \rho\right]  =\int d^{3}\mathbf{\xi}~\left\langle -\rho\left(
\mathbf{\xi}\right)  \ln\rho\left(  \mathbf{\xi}\right)  +\rho\left(
\mathbf{\xi}\right)  \right\rangle .
\end{array}
\end{align}

The problem (\ref{extreme}) leads to the following Boltzmann distribution for
the particles density $\rho$:%
\begin{equation}
\rho\left(  \mathbf{\xi}\right)  =\exp\left\langle -\mu-\beta\varphi\left(
\mathbf{\xi}\right)  \right\rangle ,
\end{equation}
where $\beta$ is the canonical parameter:%
\begin{equation}
\beta=\frac{\partial\ln\omega\left(  U\right)  }{\partial U}=\frac{3}{2\left(
U-V\right)  }, \label{canonical}%
\end{equation}
$\mu$, the chemical potential associated with the normalization condition:%
\begin{equation}
\int d^{3}\mathbf{\xi}~\rho\left(  \mathbf{\xi}\right)  =1,
\end{equation}
and $\varphi\left(  \xi\right)  $ is the gravitational potential related with
$\rho$ throughout the Poisson problem:%
\begin{equation}
\Delta\varphi\left(  \mathbf{\xi}\right)  =4\pi\rho\left(  \mathbf{\xi
}\right)  ,~\varphi\left(  1\right)  =-1,~\varphi^{\prime}\left(  1\right)
=1.
\end{equation}
The equation (\ref{canonical}) can be rewritten in order to obtain the caloric
curve $U$ \textit{versus} $\beta$:
\begin{equation}
U=\frac{3}{2\beta}+V\text{, where }V=\frac{1}{2}\int d^{3}\mathbf{\xi}%
~\rho\left(  \mathbf{\xi}\right)  \varphi\left(  \mathbf{\xi}\right)  .
\end{equation}

The solution with spherical symmetry of the above problem is numerically
implemented by introducing the function $\Phi\left(  \xi\right)  =-\mu
+\ln\beta-\beta\varphi\left(  \xi\right)  $, and solving the following
Poisson-Boltzmann problem:
\begin{equation}
\frac{1}{\xi^{2}}\frac{\partial}{\partial\xi}\left\langle \xi^{2}%
\frac{\partial\Phi\left(  \xi\right)  }{\partial\xi}\right\rangle =-4\pi
\exp\left\langle \Phi\left(  \xi\right)  \right\rangle ,
\label{poisson-boltzmann}%
\end{equation}
whose boundary conditions at the origin are:%
\begin{equation}
\Phi\left(  0\right)  =\psi,~\frac{\partial}{\partial\xi}\Phi\left(  0\right)
=0,
\end{equation}
where $\psi=\ln\left(  \beta\rho_{c}\right)  $, being $\rho_{c}$ in the
central density $\rho_{c}=\rho\left(  0\right)  $. Thus, the functions
$\rho\left(  \xi\right)  $, $\varphi\left(  \xi\right)  $ as well as the
macroscopic variables and thermodynamic potentials $U$, $\beta$, $\mu$,
$s_{0}$ are obtained as functions of the parameter $\psi$, i.e.: $\beta$ and
$\mu$ are given by:
\begin{equation}
\Phi\left(  1;\psi\right)  =-\mu+\ln\beta+\beta,~\frac{\partial}{\partial\xi
}\Phi\left(  1;\psi\right)  =-\beta.
\end{equation}
%

%TCIMACRO{\FRAME{ftFU}{3.2396in}{4.1917in}{0pt}{\Qcb{Thermostatistical
%description of the Antonov isothermal model by using the representation
%$\mathcal{R}_{U}$.}}{\Qlb{old}}{old.eps}%
%maintain-aspect-ratio TRUE;  display "ICON";  valid_file "F";
%width 3.2396in;  height 4.1917in;  depth 0pt;  original-width 6.992in;
%original-height 9.1125in;  cropleft "0";  croptop "1";  cropright "1";
%cropbottom "0";  filename 'old.eps';file-properties "XNPEU";}}}%
%BeginExpansion
\begin{figure}
[t]
\begin{center}
\includegraphics[
height=4.1917in,
width=3.2396in
]%
{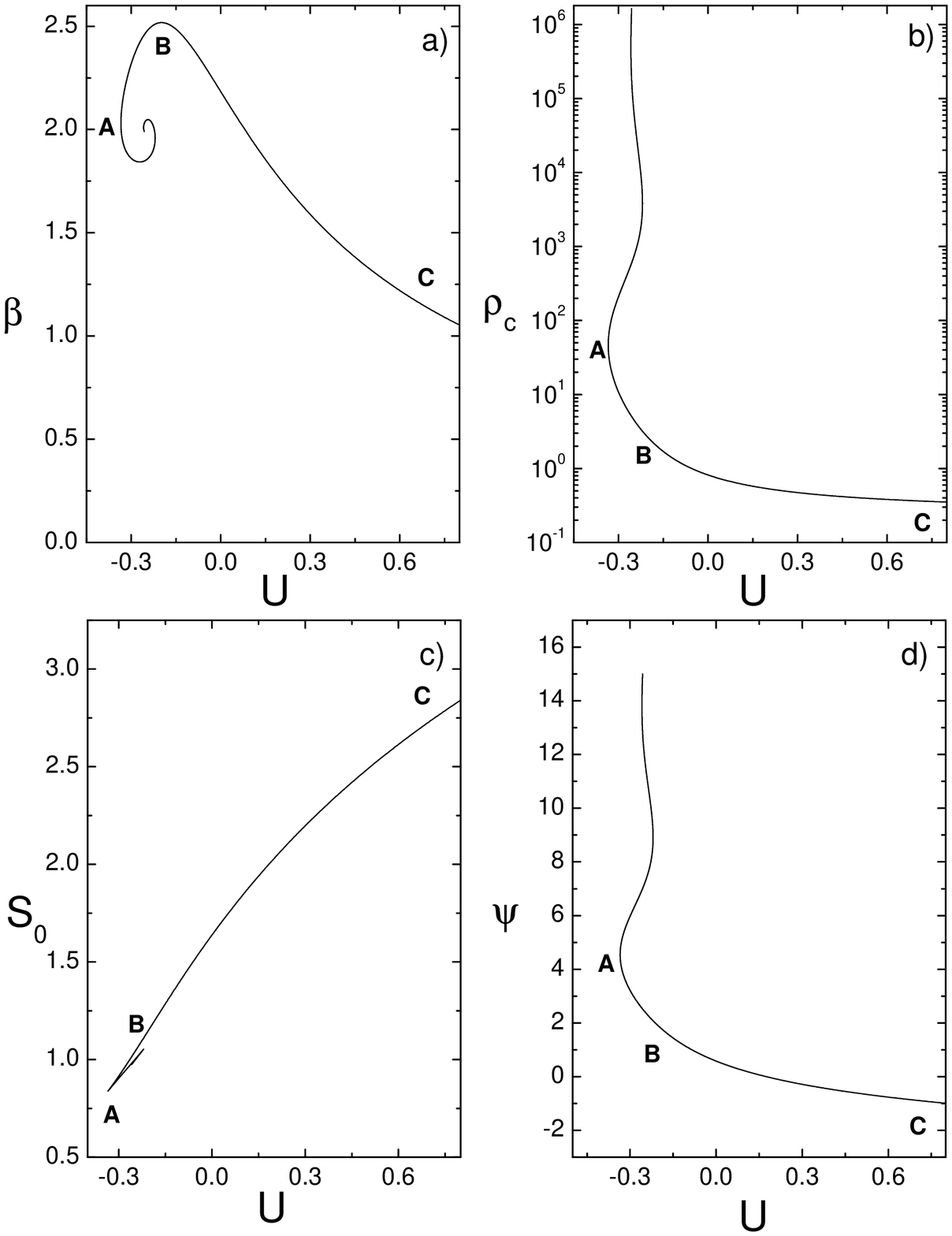}%
\caption{Thermostatistical description of the Antonov isothermal model by
using the representation $\mathcal{R}_{U}$.}%
\label{old}%
\end{center}
\end{figure}
%EndExpansion

The FIG.\ref{old} shows the microcanonical thermostatistical description of
the Antonov isothermal model. The caloric curve (panel a) and the central
density $\rho_{c}$ \textit{versus} $U$ dependence (panel b) show the existence
of thermodynamical states with a negative heat capacity (branch \textbf{A}%
-\textbf{B}), as well as the ocurrence of a gravitational collapse $\rho
_{c}\rightarrow\infty$ for low energies. The branch \textbf{A}-\textbf{B}%
-\textbf{C} corresponds to the equilibrium thermodynamic states, while the
other branch are just unstable saddle points. According to the caloric curve,
there are no equilibrium thermodynamical states when $U<U_{A}=-0.334$ or
$\beta>\beta_{B}=2.52$. The ensemble inequivalence in the branch
\textbf{A}-\textbf{B} is associated with the convexity of the entropy (panel
c) in this region.

As already discussed in our previous paper \cite{vel.self}, when this model
system initially isolate with a total energy $U\in\left(  U_{A},U_{B}\right)
$ (branch \textbf{A}-\textbf{B}) is put in a thermal contact with a heat bath
with $\beta>\beta_{B}$, it becomes unstable developing an isothermal
gravitational collapse \cite{antonov}, so that, the Antonov isothermal model
is very sensible the external influence of a heat bath in the branch
\textbf{A}-\textbf{B}. Such instability can be associated with the existence
of a first-order phase transition from a gaseous phase (with $\rho_{c}$
finite) towards a collapsed one ($\rho_{c}\rightarrow\infty$).

The above thermodynamical description was performed in the $\mathcal{R}_{U}$
representation. According to the Theorem \ref{existence}, the entropy
convexity could be avoided by considering an appropriate reparametrization
change. In order to obtain such appropriate representation, let us to observe
the $\psi$ \textit{versus} $U$ dependence also shown in panel d of the
FIG.\ref{old}. It is easy to see that, dismissing the unstable branch, there
is a biunivocal correspondence between these macroscopic observables for all
relevant equilibrium thermodynamical states. This fact supports that we can
obtain an appropriate representation $\mathcal{R}_{\varphi}$ by considering
certain experimental arrangement which keeps fixed the observable $\chi
=\exp\left(  \psi\right)  =\beta\rho_{c}$. The map $\varphi:\mathcal{R}%
_{U}\rightarrow\mathcal{R}_{\varphi}$ can be established by taking into
account that $\chi$ is the \textit{canonical parameter} in the new
representation, which is related with $\beta$ by the transformation rule
(\ref{rule_vector}):%

\begin{equation}
\chi=\exp\left(  \psi\right)  =\frac{\partial U}{\partial\varphi}%
\beta\Rightarrow\frac{d\varphi\left(  \psi\right)  }{d\psi}=\beta\left(
\psi\right)  \exp\left(  -\psi\right)  \frac{dU\left(  \psi\right)  }{d\psi}.
\label{change}%
\end{equation}

\bigskip%
%TCIMACRO{\FRAME{ftFU}{3.2396in}{4.1338in}{0pt}{\Qcb{Thermostatistical
%description of the Antonov isothermal model in the representation
%$\mathcal{R}_{\varphi}$.}}{\Qlb{new}}{new.eps}%
%maintain-aspect-ratio TRUE;  display "ICON";  valid_file "F";
%width 3.2396in;  height 4.1338in;  depth 0pt;  original-width 7.7038in;
%original-height 9.8476in;  cropleft "0";  croptop "1";  cropright "1";
%cropbottom "0";  filename 'new.eps';file-properties "XNPEU";}}}%
%BeginExpansion
\begin{figure}
[t]
\begin{center}
\includegraphics[
height=4.1338in,
width=3.2396in
]%
{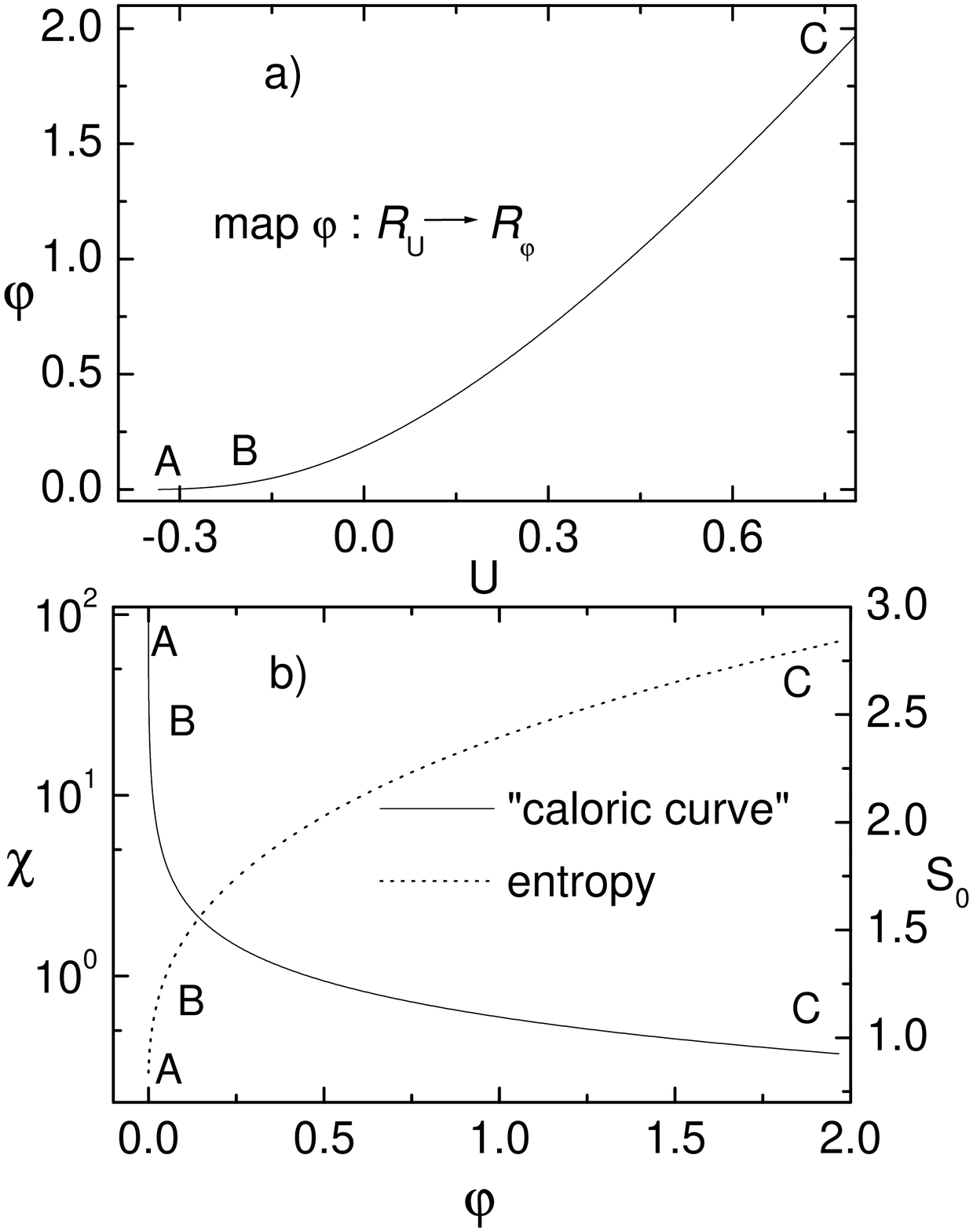}%
\caption{Thermostatistical description of the Antonov isothermal model in the
representation $\mathcal{R}_{\varphi}$.}%
\label{new}%
\end{center}
\end{figure}
%EndExpansion

FIG.\ref{new} shows the thermostatistical description of the Antonov
isothermal model in the representation $\mathcal{R}_{\varphi}$. The
reparametrization change $\varphi:\mathcal{R}_{U}\rightarrow\mathcal{R}%
_{\varphi}$ obtained from the numerical integration of the equation
(\ref{change}) is illustrated in the panel a (which is evidently a bijective
map), while the "caloric curve" $\chi$ \textit{versus} $\varphi$ and the
entropy $s_{0}$ \textit{versus} $\varphi$ dependence are shown in the panel b.
The reader can notice that no ensemble inequivalence takes place in the
representation $\mathcal{R}_{\varphi}$, and consequently, the microcanonical
description of this model system in the thermodynamic limit
becomes\ equivalent to the one carried out by considering the following
"canonical ensemble":%
\begin{equation}
\hat{\omega}_{B}\left(  \chi,N\right)  =\frac{1}{Z\left(  \chi,N\right)  }%
\exp\left\{  -\chi N\varphi\left\langle \frac{1}{E_{0}}H_{N}\right\rangle
\right\}  \label{c_like}%
\end{equation}
associated with an experimental arrangement which keeps fixed $\chi=\beta
\rho_{c}$. Since no ensemble inequivalence is observed, no topological
first-order phase transition is present in the Antonov isothermal model.

It is very important to realize that the thermostatistical description
obtained from the numerical integration of the Poisson-Boltzmann problem
(\ref{poisson-boltzmann}) makes use (in an implicit way) of the above
"canonical ensemble" since the spatial functions $\rho$ and $\varphi$, as well
as the thermodynamic variables and potentials are obtained as functions of the
parameter $\psi=\ln\chi$.

\section{Conclusions}

We have shown in the present work that an appropriate thermostatistical
description of the nonextensive Hamiltonian systems could be performed by
considering their scaling selfsimilarity properties as well as the
reparametrization invariance of the microcanonical description.

As already discussed in previous sections, the reparametrization invariance is
just an internal symmetry of the microcanonical description appearing as a
consequence of its dynamical origin. This symmetry allows us to satisfy the
necessary conditions \textbf{i)} and \textbf{ii)} (see in section \ref{ThF})
for the ensemble equivalence in the thermodynamic limit $N\rightarrow\infty$,
and the consequent well-defined character of the thermodynamic formalism based
on the Legendre transformation (\ref{Legendre}) among the thermodynamic
potentials. The consideration of the reparametrization invariance makes
possible the developing of a geometrical thermodynamic formalism in which the
ordering information based on the concavity of the Boltzmann entropy could be
defined in a topological fashion, suggesting in this way a new vision of the
phase transition concept which seems to be related with the called Topological
Hypothesis about the topological origen of the phase transitions \cite{topH2}.

Our analysis of the Antonov isothermal model shows the relevance of the
selfsimilarity and the reparametrization invariance in the context of the
astrophysical systems. An interesting remark derived from this study is that
the reparametrization invariance could be used as a powerful tool in order to
extend a canonical-like description (\ref{c_like}) for those thermodynamical
states with a negative heat capacity, fact which leads to a natural
improvement of the well-known montecarlo methods based on the consideration of
the canonical weight $\exp\left(  -\beta E\right)  $ \cite{yuki}. A very
important problem is still open: A methodology to derive the relevant scaling
selfsimilarity in a given application. Omitting some exceptional cases like
the Antonov problem reconsidered in the present paper, derivation of the
selfsimilarity scaling transformations is a very difficult task which demands
an extension of the Renormalization Group methods \cite{Gold}.

\end{document}